\documentclass[11pt,twoside]{article}
\usepackage{./asp2014}

\aspSuppressVolSlug
\resetcounters

\begin{document}

\title{Distinguishing Binary from Single L/T Transition Dwarfs Using Only Photometry}
\author{Madeline R. Wilson,$^1$ William M. J. Best,$^2$
\affil{$^1$University of California at Davis, Department of Physics, One Shields Avenue, Davis CA, USA; \email{mrwils@ucdavis.edu}}
\affil{$^2$Univeristy of Texas at Austin, Department of Astronomy, 2515 Speedway C1400, Austin, TX, USA; \email{wbest@utexas.edu}}}

\paperauthor{Madeline R. Wilson}{mrwils@ucdavis.edu}{ORCID_Or_Blank}{University of California, Davis}{Department if Physics}{Davis}{CA}{95616}{USA}
\paperauthor{William M. J. Best}{wbest@utexas.edu}{ORCID_Or_Blank}{University of Texas at Austin}{Department if Astronomy}{Austin}{TX}{78712}{USA}

\begin{abstract}
Brown dwarfs undergo little change in near-infrared luminosity while cooling from L to T spectral types, making the L/T transition (spectral types $\approx$L9--T6) an ideal region to examine the differences in photometry for binary and single brown dwarfs. We investigated the mean absolute magnitudes for binaries and singles with parallaxes as a function of J-K color in the L/T transition. Overall, binaries are on average $1-2\sigma$ brighter than singles of the same color. We discuss the implications for candidate binaries that are apparent spectral blends which span a broad range of luminosities, from the brightest binaries to the faintest singles.
\end{abstract}

\section{Background}   
\begin{figure}
\centering
\plotone{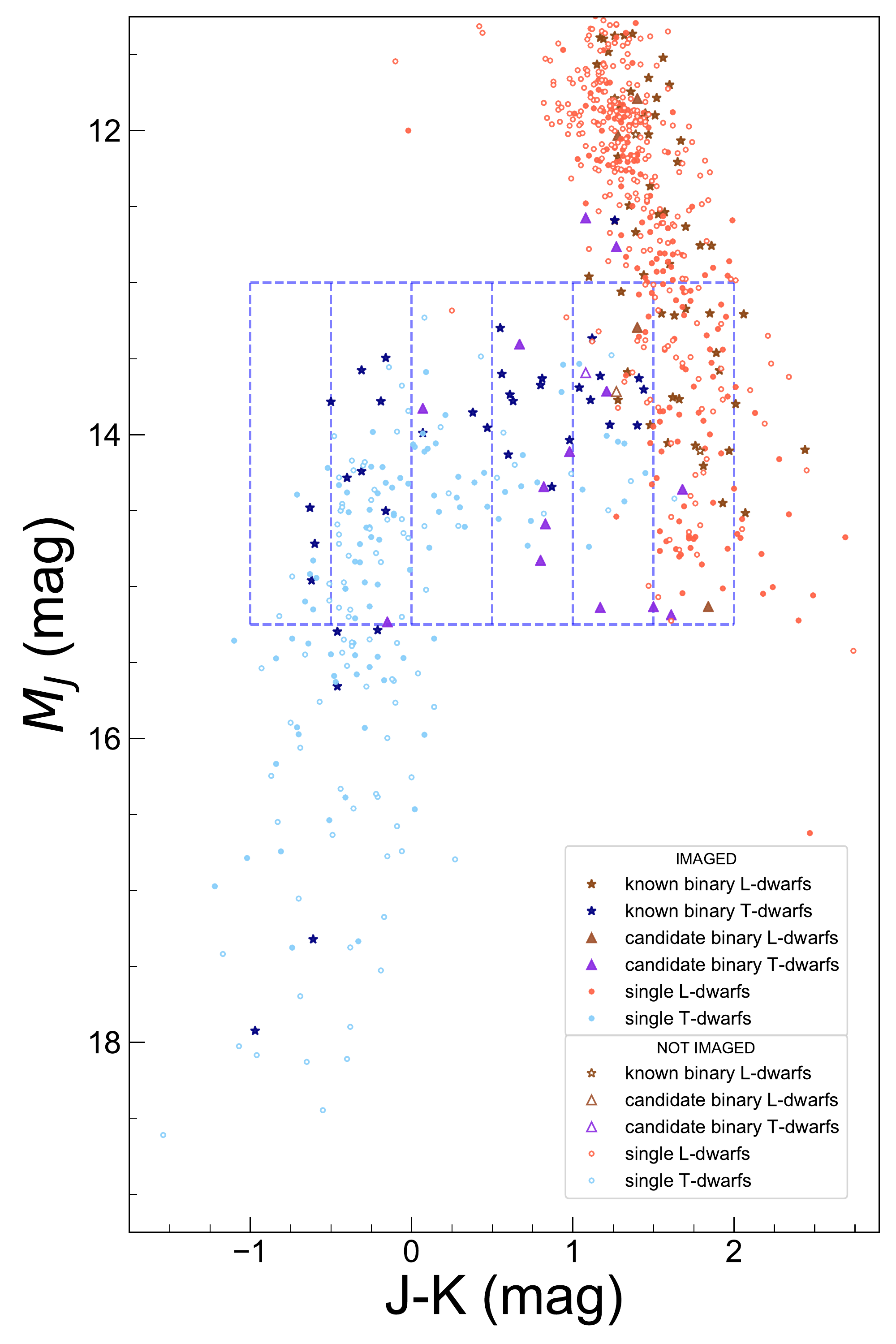}
\caption{M$_{J}$ vs. $J-K$ color-magnitude diagram for known L and T spectral type brown dwarfs. Closed and open symbols indicate objects that have and have not been observed with high angular resolution imaging, respectively. Cooling brown dwarfs evolve down the L dwarf sequence (red points at right), bluewards across the L/T transition, and then down the later-T dwarf sequence (blue points at left). The dashed blue lines framing the L/T transition define color bins described in the text and Figure~\ref{fig:2}. The L/T transition spans more than one magnitude in M$_{J}$, but is essentially a nearly-horizontal sequence for single objects.\label{fig:1}}
\end{figure}

Brown dwarfs are smaller, cooler, and fainter than main sequence stars, and they cool continuously as they age \citep[e.g.,][]{Burrows:2001iv}. As brown dwarfs transition from L to T spectral types they behave counterintuitively in the near infrared: they cool with little change in luminosity while becoming bluer \citep[e.g.,][]{Knapp:2004ji}. The L/T transition is consequently a flat region in near-infrared color-magnitude diagrams (CMDs), where we focused on the differences in photometry between single and binary brown dwarfs. 
Figure~\ref{fig:1} shows a M$_{J}$ vs. $J-K$ CMD for L and T dwarfs (Best et al., AAS Journals, submitted), in which the flat L/T transition region stands in clear contrast to the nearly vertical sequences that the early L-dwarfs and late T-dwarfs follow. 

Binaries are vital targets for measuring dynamical masses and benchmarking evolutionary models, so photometric trends distinguishing binaries from single objects can help to identify binaries that have not been resolved by high angular-resolution imaging or radial velocities.

\section{Our Sample}   
Our sample consists of all L and T dwarfs with parallaxes compiled from the Database of Ultracool Parallaxes \citep{Dupuy:2012bp,Liu:2016co}\footnote{\url{http://www.as.utexas.edu/~tdupuy/plx/Database_of_Ultracool_Parallaxes.html}}, recent literature, and Best et al. (AAS Journals, submitted). We limited our analysis to objects with parallaxes to ensure accurate luminosities. Most confirmed binaries in our sample have been resolved with high angular-resolution imaging; others were resolved in survey images (e.g., SDSS), or were confirmed via high-precision astrometry. 
Objects categorized as candidate binaries show indications of being a spectral blend  \citep[e.g.,][]{Burgasser:2010df,BardalezGagliuffi:2015fd} of two objects, but have not been confirmed by other methods. Objects not known to be binaries are categorized as singles. Some of these singles may actually be unresolved binaries with faint and/or very close companions that give no spectral indication of being a blend.

\section{Methods}   
Figure~\ref{fig:2} shows the M$_{J}$ vs. $J-K$ color-magnitude diagram of the L/T transition for our full sample as well as a volume-limited sample. Both plots include only objects for which the parallax has error~$\leq20\%$ and for which the $J-K$ color has error~$<0.2$ mag. Only objects with spectral types L9 to T6 (inclusive) were included. The volume-limited sample shown in Figure~\ref{fig:2} excludes objects with distances greater than 25~pc or outside of the declination range $-30^\circ$ to $+60^\circ$ (Best et al., in prep). We used the following parameters to define the L/T transition: $-1\leq J-K\leq2$~mag, $13\leq {\rm M}_J\leq15.5$~mag. We split the region into six $J-K$ bins of width 0.5~mag. We calculated the weighted mean M$_{J}$ and $J-K$ for the binaries, singles, and candidates in each bin, using weighted standard deviations for errors. Singles that have not been observed with high-resolution imaging were not included in calculating the mean magnitudes plotted in Figure~\ref{fig:2}.

\articlefigure[width=1.1\columnwidth]{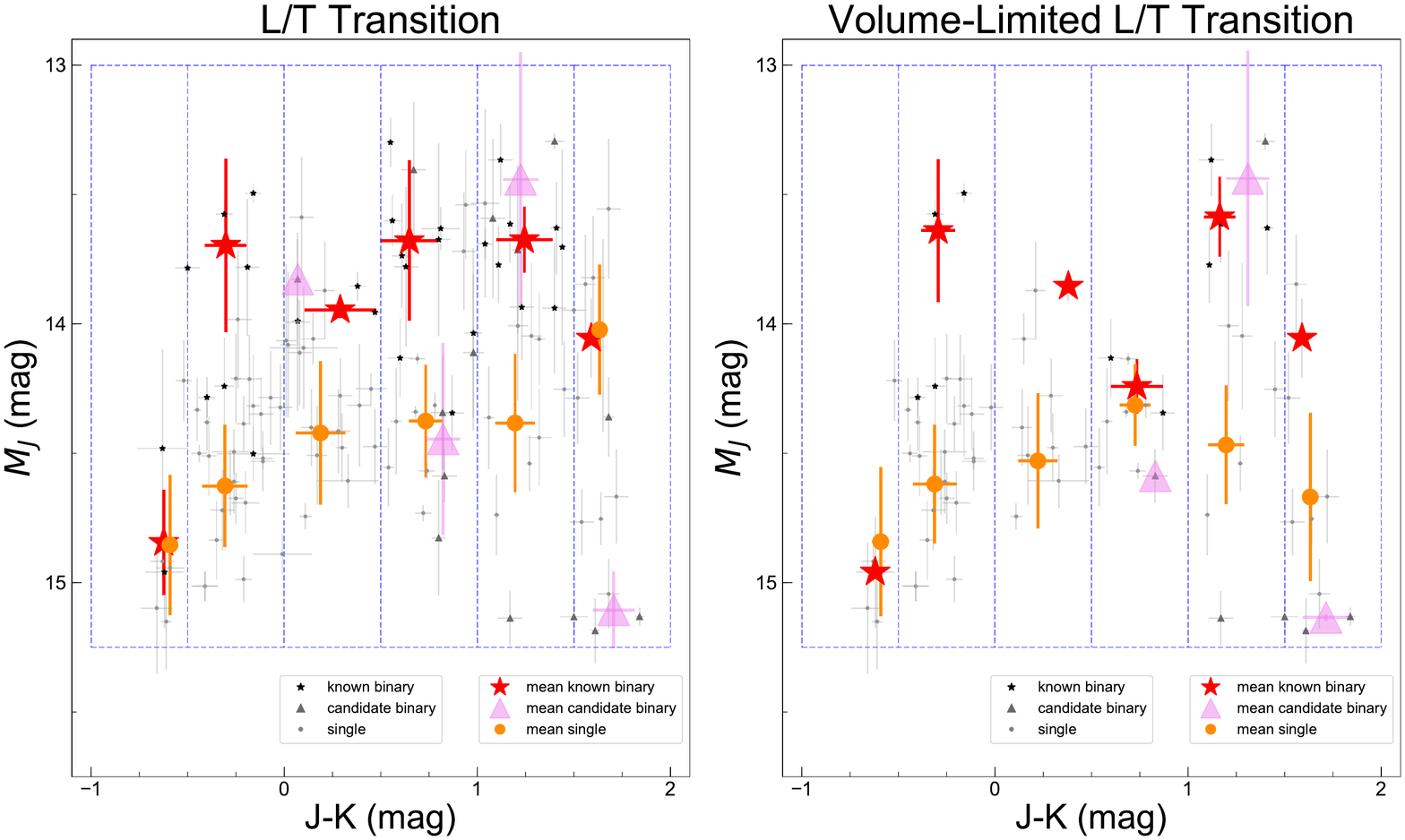}{fig:2}{{\it Left}: Color-magnitude diagram for all known L/T transition dwarfs with parallaxes, divided into six bins of 0.5~mag in $J-K$ (blue dashed lines). Single objects, candidate binaries, and confirmed binaries are represented by small gray symbols (see legend). Mean magnitudes of those types of objects are plotted in large, colored symbols for each bin. {\it Right}: Same as the left plot for a 25~pc volume-limited sample. Overall, binaries are on average $1-2\sigma$ brighter than singles of the same color. Candidate binaries span the full M$_{J}$ spread in the L/T transition, indicating that the brightest candidates are very likely binaries while the faintest are likely single objects.}

\section{Discussion}   
Overall, binaries are typically $1-2\sigma$ brighter than singles of the same color. In the $1.5\leq J-K\leq2$~mag bin of our full sample (rightmost bin in left plot of Figure~\ref{fig:2}), the mean M$_{J}$ of the singles is much brighter than in the volume-limited sample because of four bright singles that are all beyond 25~pc. It is possible that these singles are actually binaries that weren't resolved by high-resolution imaging. Because of this difference in the reddest bins, the trend for singles in the full sample looks roughly linear, whereas the trend for the singles in the volume-limited sample appears more parabolic, peaking at $J-K\approx0.5$--1~mag. The volume-limited sample likely represents the trend of the singles more accurately than the full sample, which is subject to more contamination from binaries due to Malmquist bias \citep{Malmquist:1922tn} and to the greater difficulty in resolving binaries at greater distances.

The brightest singles (M$_{J}\lesssim14$~mag) are more plausibly unidentified binaries (as they fit the overall binary trend within $1\sigma$) rather than unusually bright singles that are $\gtrsim$2$\sigma$ outliers in the population.
In addition, the candidates in the $1\leq J-K\leq2$~mag range with M$_{J}\approx15$~mag are more likely to be singles than unconfirmed binaries: in order to be so faint at those $J-K$ colors a binary would have to be comprised of two components that are both fainter than any known brown dwarfs at those colors.
These may be single objects whose transitional atmospheres feature both L~dwarf-like clouds and T~dwarf-like clear patches, giving rise to spectra that mimic blends \citep{Burgasser:2013df}.

Future work will include exploring other statistical methods for describing trends among the binaries and singles; observing the unimaged, relatively bright single objects (M$_{J}\lesssim14$~mag) with high angular-resolution imaging to determine if any are unidentified binaries; and working to confirm the candidates as either binaries or singles.

\acknowledgements We acknowledge support from the UT Austin Astronomy Department REU Program "Frontier Research and Training in Astronomy for the 21st Century" funded by NSF grant AST 1757983 (PI: Jogee) from the NSF REU program and the Department of Defense ASSURE program. WMJB received support from grant HST-GO-15238, provided by STScI and AURA.


\begin{thebibliography}{}      
\expandafter\ifx\csname natexlab\endcsname\relax\def\natexlab#1{#1}\fi
\providecommand{\url}[1]{\href{#1}{#1}}

\bibitem[{Bardalez~Gagliuffi {et~al.}(2015)Bardalez~Gagliuffi, Gelino, \&
  Burgasser}]{BardalezGagliuffi:2015fd}
Bardalez~Gagliuffi, D.~C., Gelino, C.~R., \& Burgasser, A.~J. 2015, AJ, 150,
  163

\bibitem[{Burgasser(2013)}]{Burgasser:2013df}
Burgasser, A.~J. 2013, Astronomische Nachrichten, 334, 32

\bibitem[{Burgasser {et~al.}(2010)Burgasser, Cruz, Cushing, Gelino, Looper,
  Faherty, Kirkpatrick, \& Reid}]{Burgasser:2010df}
Burgasser, A.~J., Cruz, K.~L., Cushing, M.~C., {et~al.} 2010, ApJ, 710, 1142

\bibitem[{Burrows {et~al.}(2001)Burrows, Hubbard, Lunine, \&
  Liebert}]{Burrows:2001iv}
Burrows, A.~S., Hubbard, W.~B., Lunine, J.~I., \& Liebert, J. 2001, Rev Mod
  Phys, 73, 719

\bibitem[{Dupuy \& Liu(2012)}]{Dupuy:2012bp}
Dupuy, T.~J., \& Liu, M.~C. 2012, ApJS, 201, 19

\bibitem[{Knapp {et~al.}(2004)Knapp, Leggett, Fan, Marley, Geballe, Golimowski,
  Finkbeiner, Gunn, Hennawi, Ivezi{\'c}, Lupton, Schlegel, Strauss, Tsvetanov,
  Chiu, Hoversten, Glazebrook, Zheng, Hendrickson, Williams, Uomoto, Vrba,
  Henden, Luginbuhl, Guetter, Munn, Canzian, Schneider, \&
  Brinkmann}]{Knapp:2004ji}
Knapp, G.~R., Leggett, S.~K., Fan, X., {et~al.} 2004, AJ, 127, 3553

\bibitem[{Liu {et~al.}(2016)Liu, Dupuy, \& Allers}]{Liu:2016co}
Liu, M.~C., Dupuy, T.~J., \& Allers, K.~N. 2016, ApJ, 833, 96

\bibitem[{Malmquist(1922)}]{Malmquist:1922tn}
Malmquist, K.~G. 1922, Medd. Lunds Ser. I, 100, 1

\end{thebibliography}
\end{document}